\documentstyle[preprint,aps,epsfig,eqsecnum]{revtex}
\tightenlines
\def\m@thcombine#1#2{%
  \setbox0=\hbox{$#1$}
  \setbox1=\hbox{$#2$}
  \ifdim\wd0>\wd1
    \setbox0=\hbox to\wd1{\hss\box0\hss}
  \else
    \setbox1=\hbox to\wd0{\hss\box1\hss}
  \fi
  \mathop{\vcenter{
    \offinterlineskip\box0\box1}}}
\def\lesim{\m@thcombine<\sim}
\def\gesim{\m@thcombine>\sim}
\begin{document}
 
\newdimen\picraise
\newcommand\picbox[1]
{
  \setbox0=\hbox{\input{#1}}
  \picraise=-0.5\ht0
  \advance\picraise by 0.5\dp0
  \advance\picraise by 3pt      
  \hbox{\raise\picraise \box0}
}

\draft
\title{ NONPERTURBATIVE INFRARED MULTIPLICATIVE RENORMALIZABILITY OF TWO-DIMENSIONAL
       COVARIANT GAUGE QCD}

\author{V. Gogohia, Gy. Kluge and J. Nyiri }

\address{HAS, CRIP, RMKI, Depart. Theor. Phys., Budapest 114, P.O.B. 49,
H-1525, Hungary \\
 email addresses: gogohia@rmki.kfki.hu, kluge@rmki.kfki.hu and
nyiri@rmki.kfki.hu }

\maketitle

\begin{abstract}
A nonperturbative approach to two-dimensional covariant gauge QCD
is presented in the
context of the Schwinger-Dyson equations and the corresponding Slavnov-Taylor
identities. The distribution theory, complemented by the
dimensional
regularization method, is used in order to treat correctly the
infrared singularities which inevitably appear in the theory. By
working out the multiplicative renormalization program we remove
them from the theory on a general ground and in a self-consistent
way, proving thus the infrared multiplicative renormalizability of
two-dimensional QCD within our approach. We also show explicitly how to
formulate the
bound-state problem and the Schwinger-Dyson equations for the
gluon propagator and the triple gauge field proper vertex, all free from
the infrared singularities.
\end{abstract}

\pacs{PACS numbers: 11.15.Tk, 12.38.Aw, 12.38.Lg}

\vfill

\eject

\section{Introduction}

It is well known that two-dimensional (2D) QCD is an ultraviolet (UV),
i.e., perturbativaly (PT) super-renormalizable field theory [1,2].
 However, our investigation of 2D covariant gauge QCD [3] within
the distribution theory (DT) [4] clearly shows that this theory is
infrared (IR) divergent as well since its free gluon propagator IR
singularity is nonperturbative (NP), i.e., is a severe one. For
that very reason, it becomes inevitable firstly to regularize it
and secondly to prove its IR renormalizability, i.e., to prove
that all the NP IR singularities can be removed from the theory on
a general ground and in a self-consistent way. This is just the
primary goal of this paper which is a direct continuation of our
previous work [3]. In order to formulate the IR multiplicative
renormalization (IRMR) program in 2D QCD, it is reasonable to
start with the explanation within our approach why the IR
structure of 2D QCD is so singular.

  Let us consider the Schwinger-Dyson (SD) equation for the quark
propagator (PT unrenormalized for simplicity in order not to complicate
 notations) in momentum space with Euclidean signature

\begin{equation}
S^{-1}(p) = S^{-1}_0(p) - g^2 C_F i \int {d^nq\over {(2\pi)^n}}
\Gamma_\mu(p, q) S(p-q)\gamma_\nu D^0_{\mu\nu}(q),
\end{equation}
where  $C_F$  is the eigenvalue of the quadratic Casimir
operator in the fundamental representation ( for $SU(N_c)$, in
general, $C_F = (N^2_c - 1)/2N_c = 4/3$) and
$S^{-1}_0(p) = i (\hat p + m_0)$
with $m_0$ being the current ("bare") mass of a single quark.
$\Gamma_\mu(p,q)$ is the corresponding quark-gluon proper vertex function.
Instead of the simplifications due to the limit $N_c \rightarrow \infty$
at fixed $g^2 N_c$ and light-cone gauge [5,6], we have used the
free gluon propagator in the covariant gauge from the very beginning. This
makes it possible to maintain the direct interaction of massless gluons
which is the main dynamical effect in QCD of any dimensions. In the
covariant gauge it is

\begin{equation}
D^0_{\mu\nu}(q) = i \Bigl( g_{\mu\nu} + (\xi -1) {q_\mu q_\nu \over q^2} \Bigr) {1 \over q^2},
\end{equation}
where $\xi$ is the gauge fixing parameter.

The important observation now is that for the free gluon
propagator the exact singularity $1/q^2$ at $q^2 \rightarrow 0$ in
2D QCD is severe and therefore it should be correctly treated
within the DT [4,7] (in Ref. [7] some fundamental results of pure
mathematical tractate on the DT [4] necessary for further purpose
are presented in a suitable form). In order to actually define the
system of SD equations in the IR region, it is convenient to apply
the gauge-invariant dimensional regularization (DR) method of 't
Hooft and Veltman [8] in the limit $D = 2 + 2 \epsilon, \ \epsilon
\rightarrow 0^+$. Here and below $\epsilon$ is the small IR
regularization parameter which is to be set to zero at the end of
computations. This severe singularity should be treated in the
sense of the DT (i.e., under integrals, taking into account the
smoothness properties of the corresponding test functions), where
the relation [4,7]

\begin{equation}
(q^2)^{-1} = {{\pi}\over {\epsilon}} \delta^2(q)
+ finite \ terms, \qquad \epsilon \rightarrow 0^+,
\end{equation}
holds. We point out that after introducing this expansion here and
everywhere one can fix the number of dimensions, i.e., put $D=n=2$
without any further problems since there will be no other severe
IR singularities with respect to $\epsilon$ as $\epsilon
\rightarrow 0^+$ in the corresponding SD equations but those
explicitly shown in this expansion.

Let us make a few remarks. It is worth emphasizing that the IR
singularity (1.3) is unique, the simplest IR singularity possible
in 2D QCD, however, it is a NP (severe) singularity at the same
time [4,7]. In other words, the free gluon propagator may serve as
a rather good approximation to the full gluon propagator, at least
in the deep IR region, since it exactly reproduces a possible IR
singularity of the full gluon propagator. In this connection, let
us remind that in 4D QCD the free gluon's IR singularity is not
severe, i.e., the Laurent expansion (1.3) does not exist in this
case, so it is a PT singularity there. Secondly, the DT theory
clearly shows that each skeleton diagram (which itself is the sum
of infinite series of diagrams) explicitly diverges as $1 /
\epsilon$ (see Eqs. (1.2)-(1.3)). This is true even for skeleton
diagrams containing the scattering kernels which in their turn are
defined by their own skeleton expansions (see below).

\section{IRMR program in the quark-ghost sector}

In the presence of such a severe singularity (1.3) all Green's
functions and coupling constant (which in 2D QCD has the
dimensions of mass) become generally dependent on the IR
regularization parameter $\epsilon$, i.e., they become IR
regularized (for simplicity , this dependence is not shown
explicitly in what follows). The relations between the IR
regularized and the IR renormalized (denoted with bars and which
exist as $\epsilon$ goes to zero, by definition) quantities
determine the corresponding IRMR constants. Let us remind some of
these relations introduced in Ref. [3]: $\Gamma_{\mu}(p,q) =
Z_1^{-1}(\epsilon) \bar \Gamma_{\mu}(p,q)$, $g^2 = X(\epsilon)
\bar g^2$ and $S(p) = Z_2(\epsilon) \bar S(p)$. Here and below
$Z_1 (\epsilon)$, $Z_2(\epsilon)$ and $X(\epsilon)$ are the
above-mentioned IRMR constants of the vertex, quark propagator and
coupling constant squared, respectively. The $\epsilon$-dependence
is indicated explicitly in order to distinguish them from the
usual ultraviolet (UV) renormalization constants. In all relations
containing the IRMR constants, the $\epsilon  \rightarrow 0^+$
limit ia always assumed at the final stage. The similar relations
should be introduced for the ghost degrees of freedom [3].

In their turn, the relations between these constants (established
by passing to the IR renormalized quantities in the corresponding
SD equations and Slavnov-Taylor (ST) identities) determine the
corresponding IR convergence conditions.
 The system of the corresponding IR convergence conditions in the quark-ghost
sector consists of the quark SD, the ghost
self-energy and the quark ST identity conditions. They are [3]:

\begin{eqnarray}
X(\epsilon) Z^2_2(\epsilon) Z^{-1}_1(\epsilon) &=& \epsilon , \nonumber\\
X(\epsilon) \tilde{Z}_1(\epsilon) \tilde{Z}^{-2}(\epsilon) &=& \epsilon, \qquad  \epsilon \rightarrow  0^+, \nonumber\\
Z_1^{-1}(\epsilon) \tilde{Z} (\epsilon) &=& Z_2^{-1} (\epsilon).
\end{eqnarray}
Thus, in general, we have five independent IRMR constants.
$\tilde{Z}_1(\epsilon)$  and $\tilde{Z}_2(\epsilon)$ determine the
renormalization of the ghost-gluon proper vertex and ghost
propagator, respectively, i.e., $G_{\mu} = \tilde{Z}_1(\epsilon)
\bar G_{\mu}$ and $G = \tilde{Z}_2(\epsilon) \bar G$. For
simplicity, here and below we omit the dependence on the
corresponding momenta. We know that the quark wave function IRMR
constant $Z_2(\epsilon)$ cannot be singular, while the ghost
self-energy IRMR constant $\tilde{Z}(\epsilon) = \tilde{Z}_2^{-1}
(\epsilon)$, by definition, is either singular or constant as
$\epsilon \rightarrow  0^+$. It is evident  that these very
conditions and similar ones below govern the concrete
$\epsilon$-dependence of the IRMR constants which in general
remain arbitrary.

So we have three conditions for the above-mentioned five independent IRMR
constants. Obviously, this system always has a nontrivial solution determining
three of them in terms of two any chosen independent IRMR constants.
It is convenient to choose $\tilde{Z} (\epsilon)$ and $Z_2 (\epsilon)$ as the
two independent IRMR constants since we know their possible behavior
with respect to $\epsilon$ as it goes to zero. Then a general solution to the
system (2.1) can be written down as follows:

\begin{equation}
X(\epsilon) = \epsilon Z^{-1}_2(\epsilon) \tilde{Z}(\epsilon), \quad
Z_1(\epsilon) = \tilde{Z}_1(\epsilon) = Z_2 (\epsilon) \tilde{Z} (\epsilon).
\end{equation}

Thus in the quark-ghost sector the self-consistent IRMR program
really exists. Moreover, it has room for additional
specifications. The most interesting case is the quark propagator which
is IR
finite from the very, i.e., when the
quark wave function IRMR constant $Z_2 (\epsilon) = Z_2 = const$.
The second interesting case is when
$\tilde{Z}(\epsilon) = K Z_2^{-1}(\epsilon)$, where $K$ is an arbitrary
but finite
constant (see section IV). However, in Ref. [3] it has
been shown explicitly that by redefining all the IRMR
constants, all finite
but arbitrary constants like $Z_2$ and $K$ can be put to unity without
losing generality.

\section{IR finite ST identities for pure gluon vertices}

In order to determine the IR finite bound-state problem within the
Bethe-Salpeter (BS)
formalism, it is necessary to know the IRMR constants of the three- and
four-gluon proper vertex functions which satisfy the corresponding ST
identities [1,9-13]. This information is also necessary to investigate the
IR properties of all other SD equations in 2D QCD. It is convenient to
start from the ST identity for the three-gluon vertex which is [9,10]

\begin{eqnarray}
[1 + b(k^2)] k_{\lambda} T_{\lambda\mu\nu}(k, q, r) &=& d^{-1}(q^2) G_{\lambda\nu}(q,k)
 (g_{\lambda\mu} q^2 - q_{\lambda} q_{\mu}) \nonumber\\
&+& d^{-1}(r^2) G_{\lambda\mu}(r,k) (g_{\lambda\nu} r^2 - r_{\lambda} r_{\nu}),
\end{eqnarray}
where $k+q+r=0$ is understood and $d^{-1}$ is the inverse of the exact
gluon form factor,
 while G's are the corresponding ghost-gluon vertices, and $b(k^2)$ is the
ghost self-energy. Let us now introduce
the IR renormalized triple gauge field vertex as follows:
$T_{\lambda\mu\nu}(k, q, r) = Z_3(\epsilon) \bar T_{\lambda\mu\nu}(k, q, r)$,
where $\bar T_{\lambda\mu\nu}(k, q, r)$ exists as $\epsilon$ goes to zero.
 Passing to the IR renormalized quantities, one obtains

\begin{eqnarray}
[\tilde{Z}^{-1} (\epsilon) + \bar b(k^2)] k_{\lambda} \bar T_{\lambda\mu\nu}(k, q, r)
 &=& \bar G_{\lambda\nu}(q,k) d^{-1}(q^2)(g_{\lambda\mu} q^2 - q_{\lambda} q_{\mu}) \nonumber\\
&+& \bar G_{\lambda\mu}(r,k) d^{-1}(r^2)(g_{\lambda\nu} r^2 - r_{\lambda} r_{\nu}),
\end{eqnarray}
so that the following IR convergence relation holds

\begin{equation}
 Z_3(\epsilon) = \tilde{Z}^{-1} (\epsilon) \tilde{Z}_1 (\epsilon).
\end{equation}

\begin{figure}[bp]

\[ \input{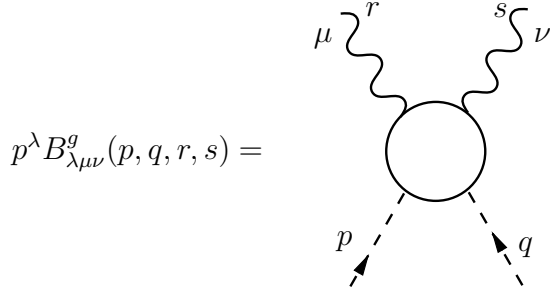} \]

\caption{ The ghost-gluon scattering kernel.}
\label{autonum}
\end{figure}

In the derivation of the relation (3.3) (as well as those similar
below) we use $b =\tilde{Z} (\epsilon) \bar b$ and $G_{\lambda\nu}
= \tilde{Z}_1(\epsilon) \bar G_{\lambda\nu}$ [3].
 Here and below we are considering the inverse of
the free gluon propagator as IR finite from the very beginning, i.e.,
$d^{-1} \equiv \bar
{d}^{-1}=1$. It is not a singularity at all and therefore it should not be
treated as a distribution [4] (there is no integration over its momentum).

The corresponding ST identity for the quartic gauge field vertex is [9,10]

\begin{eqnarray}
[1 + b(p^2)] p_{\lambda} T_{\lambda\mu\nu\delta}(p,q,r,s) &=& d^{-1}(q^2)
 (g_{\lambda\mu} q^2 - q_{\lambda} q_{\mu}) B^g_{\lambda\nu\delta}(q,p;r,s) \nonumber\\
&+& d^{-1}(r^2) (g_{\lambda\nu} r^2 -r_{\lambda}r_{\nu})
B^g_{\lambda\mu\delta}(r,p;q,s) \nonumber\\
&+& d^{-1}(s^2)(g_{\lambda\delta} s^2 - s_{\lambda} s_{\delta})B^g_{\lambda\mu\nu}(s,p;q,r)
 \nonumber\\
&-& T_{\mu\lambda\delta}(q,s,-q,-s) G_{\lambda\nu}(q+s,p,r) \nonumber\\
&-& T_{\mu\nu\lambda}(q,r,-q,-r) G_{\lambda\delta}(q+r,p,s) \nonumber\\
&-& T_{\nu\delta\lambda}(r,s,-r,-s) G_{\lambda\mu}(r+s,p,q),
\end{eqnarray}
where $p+q+r+s=0$ is understood. Here T's and G's are the
corresponding three- and ghost-gluon vertices, respectively. The
quantity $B^g$ with three Dirac indices is the corresponding
ghost-gluon scattering kernel which is shown in Fig. 1 (see also Refs. [1,11]).

Let us introduce now its IR renormalized counterpart as follows:
$B^g_{\lambda\nu\delta}(q,p; r,s) = \tilde{Z}_g(\epsilon) \bar
B^g_{\lambda\nu\delta}(q,p; r,s)$,
where $\bar B^g_{\lambda\nu\delta}(q,p; r,s)$ exists as $\epsilon$ goes to
zero. From the decomposition of the ghost-gluon proper vertex
shown in Fig. 2, it follows that
$\tilde{Z}_1(\epsilon) = (1/ \epsilon) X(\epsilon) \tilde{Z}_2 (\epsilon)
\tilde{Z}_g(\epsilon)$, so that

\begin{equation}
\tilde{Z}_g(\epsilon) = \epsilon X^{-1}(\epsilon) \tilde{Z} (\epsilon)
\tilde{Z}_1(\epsilon).
\end{equation}
Let us remind that
a factor $\sqrt{X(\epsilon)}$ should be additionally assigned to each
ghost-gluon vertex, while to the scattering kernel
$B^g$ with two gluon legs a factor $X(\epsilon)$ should be
additionally assigned.

\begin{figure}[bp]

\[ \input{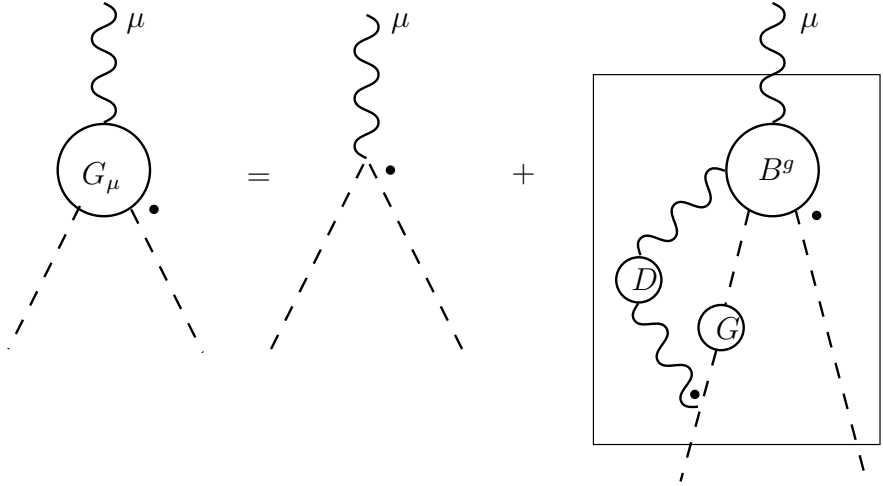}  \]

\caption{The decomposition of the ghost-gluon proper vertex. Here
and in all figures $D \rightarrow D^0$ is understood.}
\label{autonum}
\end{figure}

Let us now introduce the IR renormalized four-gluon gauge field
vertex as follows:
$T_{\lambda\mu\nu\delta}(p,q,r,s) = Z_4(\epsilon) \bar T_{\lambda\mu\nu\delta}(p,q,r,s)$,
where $\bar T_{\lambda\mu\nu\delta}(p,q,r,s)$ exists as $\epsilon$ goes to
 zero.
 Passing again to the IR renormalized quantities, one obtains

\begin{eqnarray}
[\tilde{Z}^{-1}(\epsilon) + \bar b(p^2)] p_{\lambda} \bar
T_{\lambda\mu\nu\delta}(p,q,r,s) &=& d^{-1}(q^2)
 (g_{\lambda\mu} q^2 - q_{\lambda} q_{\mu}) \bar B^g_{\lambda\nu\delta}(q,p;r,s) \nonumber\\
&+& d^{-1}(r^2) (g_{\lambda\nu} r^2 -r_{\lambda}r_{\nu})
\bar B^g_{\lambda\mu\delta}(r,p;q,s) \nonumber\\
&+& d^{-1}(s^2)(g_{\lambda\delta} s^2 - s_{\lambda} s_{\delta}) \bar
B^g_{\lambda\mu\nu}(s,p;q,r)
 \nonumber\\
&-& \bar T_{\mu\lambda\delta}(q,s,-q,-s) \bar G_{\lambda\nu}(q+s,p,r) \nonumber\\
&-& \bar T_{\mu\nu\lambda}(q,r,-q,-r) \bar G_{\lambda\delta}(q+r,p,s) \nonumber\\
&-& \bar T_{\nu\delta\lambda}(r,s,-r,-s) \bar G_{\lambda\mu}(r+s,p,q),
\end{eqnarray}
iff

\begin{equation}
 Z_4(\epsilon) = Z_3^2(\epsilon) = \tilde{Z}^{-2} (\epsilon)
 \tilde{Z}_1^2 (\epsilon).
\end{equation}
Evidently, in the derivation of this expression the general solution (2.2)
has
been  used as well as Eqs. (3.3) and (3.5). Thus we have determined the
IRMR constants of the triple and quartic gauge field vertices in Eqs. (3.3)
and (3.7), respectively.

\section{IR finite bound-state problem}

Apart from quark confinement and DBCS, the bound-state problem is one
of
the most important NP problems in QCD. The general
formalism for considering it in quantum field theory is the BS equation
(see Ref. [14] and references therein). For the color-singlet,
flavor-nonsinglet bound-state amplitudes for mesons it is shown in Figs. 3
and 4. Flavor-singlet mesons require a special treatment, since pairs,
etc.
of gluons in color-singlet states can contribute to the direct-channel
processes. The exact BS equation for the bound-state meson amplitude
$B(p,p')$ can be written analytically as follows:

\begin{equation}
S_q^{-1}(p)B(p,p')S_{\bar q}^{-1}(p') = \int d^n l K(p.p'; l) B(p.p'; l),
\end{equation}
(for simplicity all numerical factors are suppressed), where $S_q^{-1}(p)$ and $S_{\bar
q}^{-1}(p')$ are inverse quark and antiquark propagators, respectively, and $K(p,p';l)$
is the two-particle irreducible (2PI) BS scattering kernel (its skeleton
expansion is shown in
Fig. 4) which precisely defines the BS equation itself. The BS equation is
a homogeneous
linear integral equation for the $B(p,p')$ amplitude. For this reason the
meson
bound-state amplitude should always be considered as IR finite from the
very beginning, i.e., $B(p,p') \equiv \bar B(p,p')$. Passing as usual
[3] to the IR renormalized  quantities in this equation, one obtains

\begin{equation}
\bar S_q^{-1}(p)B(p,p') \bar S_{\bar q}^{-1}(p') = \int d^n l \bar K(p.p'; l)
B(p.p';l),
\end{equation}
iff

\begin{equation}
Z_2^{-2} (\epsilon) = Z_K(\epsilon), \quad \epsilon \rightarrow 0^+,
\end{equation}
where we introduce the IRMR constant $Z_K(\epsilon)$ of the BS scattering
kernel as follows: $K(p,p';  l) = Z_K(\epsilon) \bar K(p,p'; l)$ and
$\bar K(p,p'; l)$ exists as $\epsilon \rightarrow 0^+$. This
is the exact BS equation IR convergence condition.

\begin{figure}[bp]

\[ \input{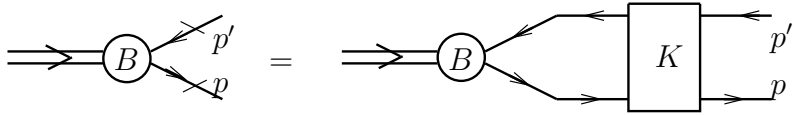}  \]

\caption{The BS equation for the flavored mesons.}
 \label{autonum}
\end{figure}

\begin{figure}[bp]

\[ \input{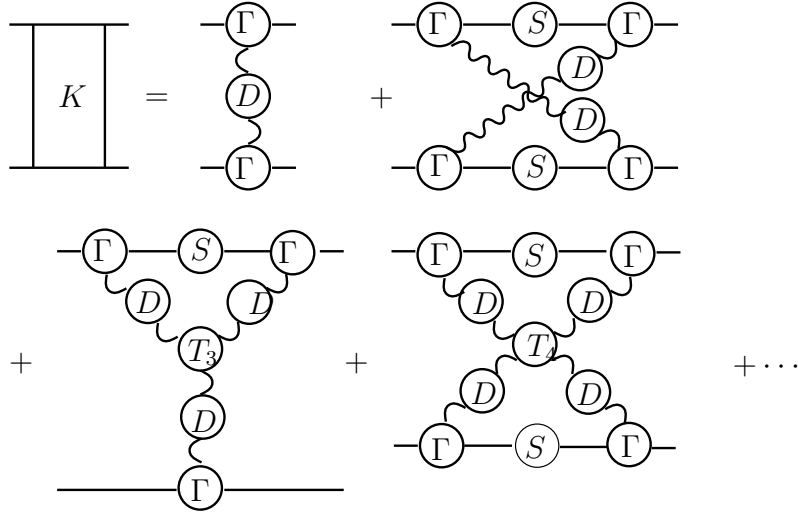}  \]

\caption{The skeleton expansion for the 2PI BS scattering kernel.}
\label{autonum}
\end{figure}

In general, the nth skeleton diagram of the BS equation skeleton expansion
contains
$n$ independent loop integrations over the gluon momentum which,
as we already know, generate
a factor $1 / \epsilon$ each, $n_1$ quark-gluon
vertex
functions and $n_2$ quark propagators. Also it contains $n_3$ and $n_4$
 three and
four-gluon vertices, respectively. It is worth reminding that to each
 quark-gluon
vertex and three-gluon vertex a factor $\sqrt{X(\epsilon)}$ should
be additionally assigned, while to the four-gluon vertex a factor
$X(\epsilon)$ should be assigned. Thus the corresponding IRMR constant
 is equal to

\begin{equation}
Z_K^{(n)}(\epsilon) = \epsilon^{-n} \Bigl[ Z_1^{-1}(\epsilon) \Bigr]^{n_1}
 \Bigl[ Z_2(\epsilon) \Bigr]^{n_2} \Bigl[ Z_3(\epsilon) \Bigr]^{n_3}
\Bigl[ Z_4(\epsilon) \Bigr]^{n_4} \Bigl[ X(\epsilon) \Bigr]^{n_4+ (n_3+n_1)/2},
 \quad \epsilon \rightarrow 0^+.
\end{equation}
On the other hand, it is easy to see that for each skeleton diagram the
following relations hold $n_2 = n_1 -2, \quad 2n = n_1 +n_3 + 2n_4$.
Substituting these relations into the previous expression as well as using the
general solution (2.2) and taking into account results of the previous
section, one finally obtains
$Z_K^{(n)}(\epsilon) =  Z_2^{-2} (\epsilon) \Bigl[ Z_2(\epsilon)
 \tilde{Z}(\epsilon) \Bigr]^{n -n_1}$, so that from Eq. (4.3) it follows
that $\Bigl[ Z_2(\epsilon) \tilde{Z}(\epsilon) \Bigr]^{n-n_1} =
A_{(n)}$, where $A_{(n)}$ is an arbitrary but finite constant
different, in principle, for each skeleton diagram. Evidently, its
solution is $ \tilde{Z}(\epsilon) = \Bigl[ A_{(n)} \Bigr]^{{-1
\over n_1 - n}} Z_2^{-1}(\epsilon)$. Let us emphasize now that the
relation between these (and all other) IRMR constants cannot
depend on which skeleton diagram is considered. This means that
the above-mentioned arbitrary but finite  constant can only be a
common factor for all skeleton diagrams, i.e., $\Bigl[ A_{(n)}
\Bigr]^{{-1 \over n_1 - n}} = K$, where $K$ can be put to unity
not losing generality as we already know. Thus the solution
becomes

\begin{equation}
\tilde{Z}(\epsilon) = Z_2^{-1}(\epsilon).
\end{equation}

Summarizing, the general system of the IR convergence conditions
for removing at this stage all the severe IR singularities on a
general ground and in a self-consistent way from the theory is

\begin{eqnarray}
X(\epsilon) &=& \epsilon Z^{-2}_2(\epsilon), \quad
\tilde{Z}^{-1}(\epsilon) = \tilde{Z}_2(\epsilon) = Z_3(\epsilon) =
\tilde{Z}_g(\epsilon) = Z_2(\epsilon), \nonumber\\
\tilde{Z}_1(\epsilon) &=& Z_1^{-1}(\epsilon) = 1, \quad Z_4(\epsilon) =
Z_3^2(\epsilon) = Z_2^2(\epsilon),
\end{eqnarray}
and the limit $\epsilon \rightarrow  0^+$ is always assumed. This
system provides the cancellation of all the severe IR
singularities in 2D QCD at this stage, and what is most important
this system provides the IR finite bound-state problem within our
approach. All the IRMR constants are expressed in terms of the
quark wave function IRMR constant $Z_2(\epsilon)$ except the
quark-gluon and ghost-gluon proper vertices IRMR constants. They
have been fixed to be unity though we were unable to investigate
the corresponding ST identity for the latter vertex [3], so we
avoided this difficulty.

\section{IR finite SD equation for the gluon propagator}

Let us now investigate the IR properties of the SD equation for
the gluon propagator which is shown diagrammatically in Fig. 5 (see
also Refs. [15,16] and references therein). Analytically it can be
written down as follows:

\begin{equation}
D^{-1}(q) = D^{-1}_0(q)  - {1 \over 2} T_t(q) -{1 \over 2} T_1(q)
- {1 \over 2} T_2(q) - {1 \over 6} T'_2(q) + T_{gh}(q) + T_q(q),
\end{equation}
where numerical factors are due to combinatorics and, for
simplicity, Dirac indices determining the tensor structure are
omitted. $T_t$ (the so-called tadpole term) and $T_1$ describe
one-loop contributions, while $T_2$ and $T'_2$ describe two-loop
contributions containing three- and four-gluon proper vertices,
respectively. Evidently, $T_{gh}, \ T_q$ describe ghost- and
quark-loop contributions.

Equating $D=D^0$ now and passing as usual to the IR renormalized
 quantities, one obtains

\begin{eqnarray}
{1 \over \epsilon} X(\epsilon) {1 \over 2} \bar T_t(q) &+& {1 \over \epsilon}
X(\epsilon) Z_3(\epsilon) {1 \over 2} \bar T_1(q) + {1 \over \epsilon^2}
X^2(\epsilon)Z_3^2(\epsilon) {1 \over 2} \bar T_2(q) + {1 \over \epsilon^2}
X^2(\epsilon)Z_4(\epsilon) {1 \over 6} \bar T'_2(q) \nonumber\\
&-& { 1 \over \epsilon} X(\epsilon) \tilde{Z}^2_2
\tilde{Z}_1(\epsilon) \bar T_{gh}(q) -  X(\epsilon)
Z_2^2(\epsilon) Z_1^{-1}(\epsilon) \bar T_q(q) = 0,
\end{eqnarray}
where the quantities with bar are, by definition, IR renormalized, i.e., they
exist as $\epsilon \rightarrow 0^+$. Let us also remind that each independent
loop integration over the gluon and ghost momenta generates a factor
$1 / \epsilon$,
while it is easy to show that there are no additional IR singularities
 with respect to $\epsilon$ in the quark loop (since we have found regular
 at zero solutions for the quark propagator [3]).

\begin{figure}

\[ \input{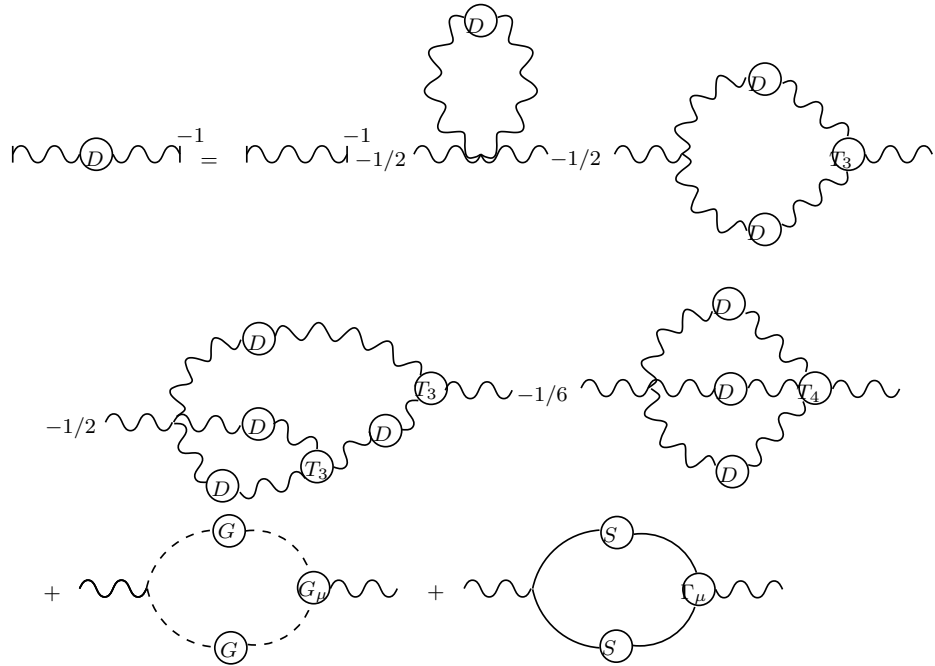} \]

\caption{The SD equation for the gluon propagator.}
\label{autonum}
\end{figure}

 Using now the general solution (4.6), one further obtains

\begin{equation}
{1 \over 2} \bar T_t(q) +Z_2(\epsilon) {1 \over 2} \bar T_1(q) +
{1 \over 2} \bar T_2(q) + {1 \over 6} \bar T'_2(q) -
Z_2^2(\epsilon) \bar T_{gh}(q) - \epsilon Z_2^2(\epsilon) \bar
T_q(q) = 0.
\end{equation}
Since the quark wave function IRMR constant $Z_2(\epsilon)$ can be only
either unity or
vanishing as $\epsilon$ goes to zero, the contribution from the
quark loop is always suppressed in the $\epsilon \rightarrow 0^+$ limit, and
we are left with
the pure Yang-Mills (YM) SD equation for the gluon propagator. For the
quark propagator which is IR
renormalized the very beginning (i.e., $Z_2(\epsilon) = Z_2 =1$, so that it
is IR finite), the SD equation (5.3) becomes

\begin{equation}
{1 \over 2} \bar T_t(q) + {1 \over 2} \bar T_1(q) + {1 \over 2}
\bar T_2(q) + {1 \over 6} \bar T'_2(q) = \bar T_{gh}(q),
\end{equation}
while for the IR vanishing type of the quark propagator ($Z_2(\epsilon) \rightarrow 0$
as $\epsilon \rightarrow 0^+$) the SD equation (5.3) becomes

\begin{equation}
\bar T_t(q) + \bar T_2(q) + {1 \over 3} \bar T'_2(q) = 0.
\end{equation}

 The tensor structure of the YM SD equations for the
gluon propagator is not important here. However, it may simplify
substantially the corresponding IR renormalized YM SD equations
(5.4) and (5.5). What matters here is that within our approach the
self-consistent equations for the gluon propagator free from the
severe IR singularities exist in the YM sector.
 In other words, the general
solution (4.6) eliminates all the severe IR singularities from Eq.
(5.1), indeed.

\section{IR finite SD equation for the three-gluon proper vertex}

It is instructive to investigate the IR properties of the SD equation for the
triple gauge field proper vertex since it provides a golden opportunity
to fix $Z_2(\epsilon)$. This equation is shown in Fig. 6. For
simplicity, the skeleton
expansions of the corresponding kernels $M, \bar M'$ and $B^g$ are not
shown. Let us note that the ghost-gluon scattering kernel
$B^g$ (for which we have already established its IRMR constant from the
decomposition of the ghost-gluon proper vertex shown in Fig.2) is denoted
as $G'$ in Ref. [1].

\begin{figure}[bp]

\[ \input{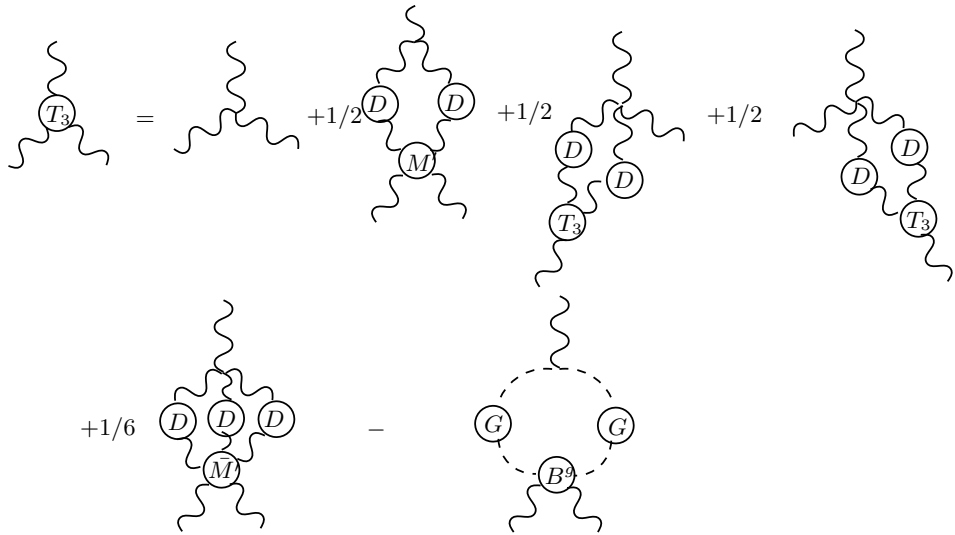}  \]

\caption{The SD equation for the triple gauge field vertex.}
\label{autonum}
\end{figure}

 Obviously, there is no need
to investigate separately the IR properties of the SD equations
for the quark-gluon vertex and for pure gluon vertices since the
information about their IRMR constants has been uniquely extracted
from the corresponding ST identities (see the general solution
(4.6)). Moreover, the IRMR constants of different types of the
scattering kernels which enter the above-mentioned SD equations
(see, for example, Fig. 6) are to be precisely determined by the
general system (4.6). In principle, each skeleton diagram of the
above-mentioned expansions can be investigated in the same way as
the BS scattering kernel was investigated in section IV.

The above-mentioned golden opportunity is provided by the third
and fourth terms of this SD equation. The interesting feature of
these terms is that they do not contain unknown scattering
kernels, so their IR properties can be investigated directly by
using only the known IRMR constants. On the other hand, these
terms are nothing but the corresponding independent decompositions
of the triple gauge field proper vertex with the IRMR constant is
equal to $Z_3(\epsilon) = Z_2(\epsilon)$, so one has

\begin{equation}
Z_2(\epsilon) = {1 \over \epsilon} X(\epsilon)
Z_2(\epsilon) = Z_2^{-1}(\epsilon) = 1,
\end{equation}
which, obviously, has only a unique solution given by the last equality.
Thus we
have fixed finally the quark wave function IRMR constant to be unity.
Adopting the same method, it is easy to show that all other IRMR constants for
the corresponding scattering kernels are:

\begin{equation}
\tilde{Z}_g(\epsilon) = Z_{M'}(\epsilon) = Z_{\bar M'}(\epsilon) =1.
\end{equation}

We are now ready to investigate the IR properties of the SD
equation for the triple gauge field proper vertex shown in Fig. 6
without refereing to the skeleton expansions of the corresponding
scattering kernels. Using the previous results, the
IR renormalized version of this equation is

\begin{equation}
\bar T_3 = T_3^{(0)} + {1 \over 2} \bar T_1
+ {1 \over 2} \bar T'_1 + {1 \over 2} \bar T^{''}_1 + {1 \over 6} \bar T_2  -
\bar T_g,
\end{equation}
where, for simplicity, we omit the dependence on momenta and
suppress the Dirac indices (i.e., tensor structure). As usual, the quantities
with bar are, by definition, IR renormalized, i.e., they
exist as $\epsilon \rightarrow 0^+$.

The SD equations for all other Green's functions can be
investigated in the same way, in particular for the quark-gluon
proper vertex shown for example in Ref. [1]. The general solution
(4.6), taking into account the fundamental relation (6.1),
provides their IR convergence, i.e., they exist in the $\epsilon
\rightarrow 0^+$ limit and, hence, similar to the SD equations,
explicitly considered here, they are free of the severe IR
divergences with respect to $\epsilon$. In particular, let us note
that Eq. (5.5) should be ruled out as a possible SD equation for
the gluon propagator and the SD equation (5.4) which includes the
ghost loop is the only possible one.

\section{Discussion and Conclusions}

In summary, the main observation is that 2D QCD is an inevitably
IR divergent theory, and therefore a few points are worth
reemphasizing.

 We have shown explicitly how the NP IRMR program should be done to
remove all the severe IR singularities from the theory on a
general ground and in a self-consistent way. The general system of
the IR convergence conditions (4.6), on account of the fundamental
relation (6.1), simply becomes

\begin{equation}
X(\epsilon) = \epsilon, \quad \epsilon \rightarrow  0^+,
\end{equation}
while all other independent quantities (Green's functions) are IR
finite from the very beginning, i.e., their IRMR constants are
simply unity. Evidently, only the nontrivial IR renormalization of
the coupling constant squared is needed to render the theory IR
finite, i.e., to make it free from all the severe IR divergences.
Only the condition (7.1) provides a cancellation of all the severe
IR singularities in 2D QCD. This completes the proof of the NP IR
multiplicative renormalizability of 2D QCD within our approach.

 Our proof implies that quark propagator should be
IR finite from the very beginning, i.e., $Z_2(\epsilon)=1$ which
means $S(p) = \bar S(p)$. In the 't Hooft model [5], the quark
propagator is IR vanishing, i.e., $Z_2(\epsilon)$ goes to zero as
$\epsilon \rightarrow 0^+$.  However, there is no contradiction
with the above-mentioned since in his model it is implicitly
assumed that neither $g^2$ nor $N_c$ depend on $\epsilon$ (i.e.,
they are IR finite from the very beginning) though both parameters
$g^2$ and $N_c$ (since it is free one in the large $N_c$ limit
approach) should, in principle, depend on it in the presence of
such a severe IR singularity in the theory. From our general
solution (4.6) then it follows that $Z_2(\epsilon) =
\sqrt{\epsilon}$, indeed, since in this case one has to put
$X(\epsilon)=1$, i.e., the coupling constant squared is IR finite
from the very beginning ($g^2 = \bar g^2$).

A few remarks are in order. In principle, no regularization scheme
(how to introduce the IR regularization parameter in order to
parameterize the severe IR divergences) should be introduced "by
hand". First of all it should be well defined. Secondly, it should
be compatible with the DT [4]. The DR scheme [8] is well defined;
in Ref. [7] we have shown how it should be introduced into the DT
(complemented by the number of subtractions, if necessary). In
principle, other regularizations schemes are also available, such
as, e.g., analytical regularization used in Ref. [17] or the
so-called Speer's regularization [18]. However, they should be
compatible with the DT as was emphasized above. Anyway, not the
regularization is important but the DT itself.

 Whether the theory
is IR multiplicative renormalizable or not depends on neither the
regularization nor the gauge. Due to the chosen regularization
scheme or the gauge only the details of the corresponding IRMR
program can be simplified. For example, in the light-cone gauge at
any chosen regularization scheme (the 't Hooft model with
different prescriptions how to deal with the severe IR
singularities [2] (and references therein)) to prove the IR
multiplicative renormalizability of 2D QCD is almost trivial. This
is mainly due to the fact that in this case only two sectors
survive in QCD, namely quark and BS sectors. In other words, if
theory is proven to be IR or UV renormalizable in one gauge, it is
IR or UV renormalizable in any other gauge. This is true for the
regularization schemes as well. As it follows from the present
investigation, to prove the IR multiplicative renormalizability of
2D QCD in the covariant gauge was not so simple. However, it was
necessary to get firstly the IR finite bound-state problem (which
is important for physical applications), and secondly to
generalize our approach on 4D QCD which is real theory of strong
interactions. 2D QCD in the light-cone gauge is not appropriate
theory for this purpose since its confinement mechanism looks more
like that of the Schwinger model [1] of 2D electrodynamics, than
it may happen in real QCD, where we believe it is much more
complicated.

 The structure of the severe IR singularities in Euclidean space
is much simpler than in Minkowski space, where kinematical
(unphysical) singularities due to light cone also exist. In this
case it is rather difficult to correctly untangle them from the
dynamical singularities, only ones which are important for the
calculation of any physical observable. Also the consideration is
much more complicated in configuration space [4]. That is why we
always prefer to work in momentum space (where propagators do not
depend explicitly on the number of dimensions)  with Euclidean
signature. We also prefer to work in the covariant gauges in order
to avoid peculiarities of the noncovariant gauges [19], for
example how to untangle the gauge pole from the dynamical one. The
IR structure of 2D QCD in the light-cone gauge by evaluating
different physical quantities has been investigated in more detail
in Refs. [2,20-23] (and references therein).

The only dynamical mechanism which can be thought of in 2D QCD is
the direct interaction of massless gluons (without explicitly
involving some extra degrees of freedom). It is well known that in
the deep UV limit this interaction brings to birth asymptotic
freedom (AF) [1] in QCD as well. It becomes strongly singular in
the deep IR domain and can be effectively correctly absorbed
(accumulated) into the gluon propagator.  Let us remind that
formally the $D=D^0$ solution always exists in the system of the
SD equations due to its construction by expansion around the free
field vacuum. Otherwise it would be impossible to introduce
interaction step by step though the final equations make no
reference to perturbation theory [1]. It is either trivial
(coupling is zero) or nontrivial, then some additional
condition(s) (constraint(s)), involving other Green's functions,
is to be derived. Eq. (5.4) is just this exact constraint. The
only question to be asked is whether this solution (precisely to
its severe IR structure) is justified to use in order to explain
some physical phenomena, such as quark confinement, DBCS, etc. or
not. By proving the NP IR multiplicative renormalizability of 2D
QCD, we conclude that this is so, indeed. At the same time, let us
emphasize that this solution is not justified to use for the
above-mentioned purposes in 4D QCD since its IR singularity is not
severe (i.e., NP) there as was mentioned in the Introduction.

One can also conclude that in some sense it is easier to prove the
IR multiplicative  renormalizability of 2D QCD than to prove its
UV renormalizability. The reason is, of course, that we know the
mathematical theory which has to be used - the theory of
distributions [4]. This is due to its fundamental result [4,7]
which requires that any NP (severe) singularity with respect to
the momentum in the deep IR domain in terms of $\epsilon$ should
be always $1 / \epsilon$ and this does not depend on how the IR
regularization parameter $\epsilon$ has been introduced in the way
compatible with the DT itself. On the other hand, the
above-mentioned fundamental result relates the IR regularization
to the number of space-time dimensions [4,7] (compactification).
It is clear that otherwise none of the IRMR programs would be
possible. In other words, the DT provides the basis for the
adequate mathematical investigation of the global character of the
severe IR divergences (each independent skeleton loop diagram
diverges as $1/ \epsilon$), while the UV divergences have a local
character, and thus should be investigated term by term in powers
of the coupling constant.

In this connection let us make a few remarks. The full dynamical
content of 2D (4D) QCD is contained in its system of the SD
equations of motion. To solve 2D (4D) QCD means to solve this
system and vice versa. In particular, to prove the IR
renormalizability of 2D (4D) QCD means to formulate the IRMR
program for removing all the NP IR singularities from this system
on a general ground and in a self-consistent way. As was mentioned
above, the fortunate feature which makes this possible is the
global character of the IR singularities in 2D QCD. Each skeleton
diagram is the sum of infinite series of terms, however, the DT
shows how their severe IR singularities can be summed up.
Moreover, it shows how the severe IR singularities of different
scattering kernels (which by themselves are infinite series of the
skeleton diagrams) can also be summed up.

The next important step is to impose a number of independent
conditions to cancel all the NP IR singularities which inevitably
appear in the theory after the above-mentioned summations have
been done with the help of the entire chain of strongly coupled SD
and BS equations. They should also be complemented by the
corresponding ST identities which are consequences of the exact
gauge invariance, and therefore are exact constraints on any
solution to QCD [1]. The only problem now is to find
self-consistent solutions to the system of the IR convergence
conditions. If such solutions exist, so everything is O.K. If not,
the theory is not IR renormalizable. It is worth reemphasizing
that we have found a self-consistent solution to this system (Eq.
(7.1)).

Concluding, we would like to make a few things perfectly clear. In
Ref. [3] we made sharp conclusions that 2D covariant gauge QCD
implies quark confinement and dynamical breakdown of chiral
symmetry. However, these conclusions would be groundless if in the
present investigation we were not be able to prove that 2D QCD is
an IR multiplicative renormalizable theory. This is our first
conclusion here, and it is based on the compelling mathematical
ground provided by the DT itself.

The second one is that we fixed finally the type of the quark
propagator. The NP IRMR program implies it to be IR finite from
the very beginning.

The third conclusion (the most important for physical
applications) is that the bound-state problem becomes tractable
within our approach. To any order in the skeleton expansion of the
BS scattering kernel shown in Fig. 7, the corresponding BS
equation (4.2) can be reduced finally to an algebraic problem. Its
derivation is absolutely similar to the derivation of the IR
renormalized quark SD equation which has been carried out in Ref.
[3] (see also Ref. [24], where one can find a more detail
discussion, conclusions and the comparison with the 't Hooft model
as well).

It was widely believed that the severe IR singularities could not
be put under control. However, we show explicitly here and in Ref.
[3] that the above-mentioned common belief is not justified. They
can be controlled in all sectors of QCD of any dimensions by using
correctly the DT [4,7]. This can be considered also as one of our
important results from a mathematical point of view.

 This work has been carried out
under BPCL-17 Contract and we thank L.P. Csernai for support. One
of the authors (V.G.) is grateful to A.V. Kouziouchine for help
and support.

 \vfill

\eject

\end{document}